\documentclass[11pt, draftclsnofoot, onecolumn]{IEEEtran}
\ifCLASSINFOpdf

\else

\fi
\usepackage[cmex10]{amsmath}
\usepackage{amssymb}
\usepackage{cite}
\usepackage{graphicx}
\usepackage{array,color}
\usepackage{amsmath}
\usepackage{stfloats}
\usepackage{graphicx}
\usepackage{subfigure}
\usepackage{tabularx}
\usepackage{epsfig,epsf,color,balance,cite}
\usepackage{algorithmic}
\usepackage{algorithm}
\usepackage{bm}
\usepackage{textcomp}

\begin{document}

\title{User-centric C-RAN Architecture for  Ultra-dense 5G Networks:  Challenges and Methodologies}
\author{ Cunhua Pan, \IEEEmembership{Member, IEEE},  Maged Elkashlan, \IEEEmembership{Member, IEEE}, Jiangzhou Wang, \IEEEmembership{Fellow, IEEE}, Jinhong Yuan, \IEEEmembership{Fellow, IEEE}, and Lajos Hanzo, \IEEEmembership{Fellow, IEEE}
\thanks{\emph{Cunhua Pan and Maged Elkashlan are with the Queen Mary University of London, London E1 4NS, U.K. (Email:\{c.pan, maged.elkashlan\}@qmul.ac.uk).}}
\thanks{\emph{Jiangzhou Wang is with the School of Engineering and Digital Arts, University of Kent, Canterbury, Kent, CT2 7NZ, U.K. (e-mail: j.z.wang@kent.ac.uk).}}
\thanks{ \emph{Jinhong Yuan is with the University of New South Wales, Sydney, NSW 2052, Australia. (e-mail:j.yuang@unsw.edu.au).}}
\thanks{\emph{Lajos Hanzo is with the School of Electronics and Computer Science, University of Southampton, Southampton, SO17 1BJ, U.K. (e-mail:lh@ecs.soton.ac.uk). }}
}


\maketitle
\vspace{-1.9cm}
\begin{abstract}
Ultra-dense networks (UDN) constitute one of the most promising techniques of supporting the 5G mobile system. By deploying more small cells in a fixed area, the average distance between users and access points can be significantly reduced, hence a dense spatial frequency reuse can be exploited. However, severe interference  is the major obstacle in UDN. Most of the contributions deal with the interference by relying on cooperative game theory. This paper advocates the application of dense user-centric C-RAN philosophy to UDN, thanks to the recent development of cloud computing techniques. Under dense C-RAN, centralized signal processing can be invoked for supporting CoMP transmission. We summarize the main challenges in dense user-centric  C-RANs. One of the most challenging issues is the requirement of the global CSI for the sake of cooperative transmission. We investigate this requirement by only relying on partial CSI, namely, on inter-cluster large-scale CSI. Furthermore, the  estimation of the intra-cluster CSI is considered, including the  pilot allocation and robust transmission. Finally, we highlight several promising research directions to make the dense user-centric  C-RAN become a reality, with special emphasis on the application of the `big data' techniques.

\end{abstract}

\IEEEpeerreviewmaketitle
\section{Introduction}
The fifth generation (5G) wireless system is anticipated to offer a substantially  increased data throughput compared to the fourth generation (4G) system. To achieve this ambitious goal, ultra dense networks (UDN) have been widely regarded as one of the most promising solutions \cite{Andrews2014}. In UDN, the average distance between users and small cell BSs is significantly reduced, hence the link quality is dramatically improved, which further increases the network capacity.

However, the drastic interference generated by the neighboring small cells is a limiting factor in UDN. The attainable network performance may even be decreased when the BS density is extremely high \cite{MingDing2016}. Hence, the interference should be carefully managed in order to reap the potential benefits of UDN. Most of the existing contributions deal with the interference by designing partially distributed algorithms based on game-theoretical approaches.  By adopting cooperative game theory, multiple small cell BSs exchange the necessary information for their coordination through the wired backhaul (BH) links, which works well for small-scale networks. However, for UDN, the heavy overhead of coordination and the increasing cost of deploying the wired BH links will preclude the application of this approach. Apart from the interference issues, employing more small cell BSs will also increase the maintenance and operational costs.  Hence, a new network architecture should be adopted to support  reliable communications in UDN.

\begin{figure}
\centering
\includegraphics[width=6in]{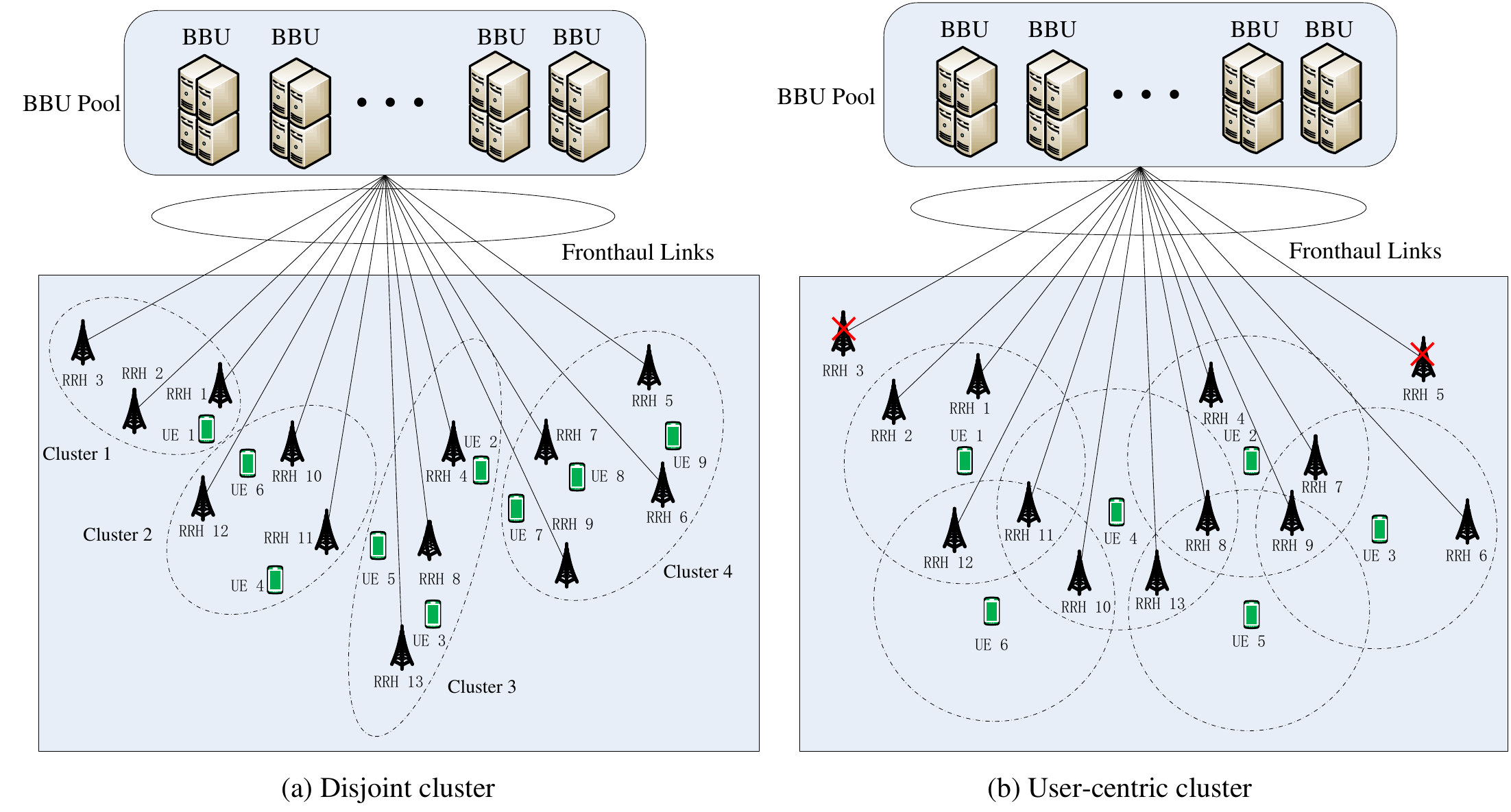}
\caption{Illustration of a C-RAN  architecture: (a) Disjoint cluster, where the whole network is divided into several non-overlapped clusters, and the cluster-edge users still suffer from high inter-cluster interference; (b) User-centric cluster, where cluster is formed from the user side and the cluster-edge issues are eliminated. }
\label{fig1}
\end{figure}

Due to the recent developments in cloud computing, the ultra-dense cloud radio access network (C-RAN) concept has been regarded as a promising network architecture that can efficiently address the issues arising in UDN.

The ultra-dense C-RAN architecture is shown in Fig.~\ref{fig1}, which consists of three key components:
\begin{enumerate}
  \item Baseband unit (BBU) pool supported by the techniques of cloud computing, network function virtualization (NFV),  software-defined networks (SDN), and so on.
  \item Low-cost, low power radio remote heads (RRHs) distributed over the coverage area.
  \item Wireless fronthaul links that connect the RRHs to the BBU pool.
\end{enumerate}

The main characteristic of dense C-RAN is that all the baseband signal processing units of conventional small cell BSs have been incorporated in the BBU pool, where the computing resources can be shared among the BBUs. Then  conventional full-functionality small cell BSs can be replaced by the low-cost, low-complexity RRHs, which are only responsible for the transmission/reception. Hence,  RRHs can be densely deployed at a low hardware cost to provide ultra-high throughput and seamless coverage for a large number of users in tele-traffic hot spots, such as airports and shopping malls. Thanks to the centralized architecture of C-RAN, the global network information can be shared in the BBU pool and  cooperative communication techniques can be realized with the aid of powerful cloud computing, such as large-scale network coordination, global resource management, coordinated multi-point (CoMP) processing, etc.  Hence, the cochannel interference that is a limiting factor in UDN can be efficiently mitigated under the ultra-dense C-RAN architecture. Additionally, the operating status of RRHs and the computing resources of the BBU pool can be dynamically controlled in order to adapt to the capacity demand fluctuations of the users, which leads to significant energy and operational cost reductions. In \cite{Tang2017}, Tang \emph{et al.} showed that the system cost, which consists of computation cost and the wireless transmission cost (including both transmit power and circuit power), can be reduced by roughly twelve percent  through jointly optimizing the number of active virtual machines (VMs) and operating status of RRHs and fronthaul links. Table \ref{tab1} contrasts the differences between conventional small cell network and the ultra-dense C-RAN.
\begin{table}[!t]
\renewcommand{\arraystretch}{1.1}
\caption{Different Types of UDN Deployment}
\label{tab1}
\centering
\begin{tabular}{|c|c|c|c|c|}
\hline
 \textbf{Type of UDN} &\textbf{Functionality} &\textbf{Interference Management}& \textbf{Connectivity} & \textbf{Deployment Cost}\\
\hline
Small cells& Fully functioning& Distributed algorithm & Wired backhaul & High maintenance cost\\
\hline
Dense C-RAN &  PHY layer functioning& Centralized algorithm & Wireless fronthaul& Low hardware cost\\
\hline
\end{tabular}
\end{table}

However, there are some challenges that need to be handled in dense C-RAN before its successful practical deployment. The contributions of this paper are summarized as follows:
\begin{enumerate}
  \item We first summarize the unique research challenges in dense C-RAN and the existing solutions for each challenge. We place an emphasis on the challenge of heavy training overhead for estimating all the channel state information (CSI) for CoMP transmission. Most of the existing works consider perfect intra-cluster CSI, which is impractical due to the limited amount of pilot resources.
  \item We provide a complete framework to deal with the imperfect intra-cluster CSI. One novel low-complexity pilot allocation algorithm is first proposed by considering the multi-user pilot interference. Then, robust beamforming vector design is considered to account for the CSI estimation error.
  \item Some promising research directions are highlighted, especially the cluster design by the aid of `Big Data' technique.
\end{enumerate}

The rest of this paper is organized as follows: in Section \ref{state}, the new challenges for dense C-RAN are provided; Section \ref{robust} provides a complete procedure to deal with the imperfect CSI, including both the pilot allocation and robust beamforming design. Simulation results are also shown in Section \ref{robust}. Finally, conclusions are drawn in Section \ref{hrforh} along with some future research directions.

\section{Research Challenges and State of the Art Solutions}\label{state}
In this section, we summarize the challenges arising in dense C-RAN along with the state of the art solutions.

\textbf{High Computational Complexity}: In dense C-RAN, the BBU pool usually supports a large number of RRHs and the number of variables to optimize, such as beamforming-vectors will become excessive, even in the context of cloud computing. The most common technique of reducing the complexity is to adopt the cluster technique. In general, there are two types of cluster techniques, as shown in Fig.~\ref{fig1}: Disjoint clustering and user-centric clustering. In the disjoint cluster, all the RRHs in the network are partitioned into several non-overlapped clusters, and the RRHs in each cluster employ the CoMP technique to serve the users within the coverage area of this cluster. However, the cluster-edge users still suffer from severe inter-cluster interference.  By contrast, in the user-centric cluster, each user is individually served by its nearby RRHs. The scheduled user is the center of the cluster. Different clusters may overlap with each other, which will eliminate the potential cluster-edge effect.  Hence, the user-centric cluster constitutes the focus of this paper.

\textbf{Stringent Fronthaul Capacity Requirement}: In conventional C-RAN, the fronthaul links are typically wired links, such as optical fibers. However, in ultra-dense C-RAN a large number of fronthaul links are required. Laying wired links requires high operational and maintenance costs. An attractive alternative is to use wireless fronthaul links, such as millimeter wave (mmWave) Communications, which are much more scalable and cost-effective than wired links. However, the bandwidth of wireless links is much lower than that of wired links, which means that the number of users supported by each wireless link is much lower. The fronthaul capacity constraint has been extensively studied, which can be divided into two categories: the compression strategy and the data sharing strategy. In the compression strategy, the BBU pool first computes the beamforming-vectors for each RRH. Then, the beamformed signals are generated at the BBU pool, which are compressed and sent to the corresponding RRHs. The fronthaul capacity is related to how fine is the resolution of the compressed signals: higher resolutions require a higher fronthaul capacity. Hence, the compression resolution should be optimized under the fronthaul capacity constraints. In \cite{Park2013}, Park \emph{et al.} proposed to leverage joint compression of the signals of different RRHs to better optimize the compression resolutions. In the second strategy, the beamforming-vectors computed at the BBU pool are directly sent to the corresponding RRHs. Then, the BBU pool shares each user's data directly with its serving cluster. The beamformed signals are generated at each RRH. In this strategy, the fronthaul capacity depends on the number of users served by each link.  Hence, the user-RRH associations should be optimized under the fronthaul capacity constraints. In  \cite{Dai2014}, Dai \emph{et al.} considered the network utility maximization problem for the user-centric downlink C-RAN, where the per-RRH fronthaul capacity constraints are explicitly taken into account.

\textbf{Huge Training Overhead for CSI Estimation}: To facilitate CoMP transmission, the global CSI should be available at the BBU pool, which constitutes an excessive channel estimation overhead for dense C-RAN.  Caire \emph{et al.} \cite{Caire2010} showed that the increasing amount of training overhead may  outweigh the gains provided by CoMP.
To reduce the overhead, a promising technique is to rely on partial CSI case under the user-centric cluster, where each user only estimates the CSI of the links from the RRHs within its cluster (termed as intra-cluster CSI) and only tracks the path loss and shadowing  outside its own cluster (termed as inter-cluster CSI), which are sent back to the BBU pool. These parameters are necessary for the CoMP design at the BBU pool. Indeed, the large-scale fading may be readily tracked since it changes slowly compared to the instantaneous CSI. The design of beamforming vectors under this partial CSI case is a challenging task. Hence, there is  a paucity of contributions \cite{Shi2014ICC}-\cite{pan2017jsac} based on partial CSI. To elaborate,  compressive CSI acquisition was proposed in \cite{Shi2014ICC} for determining the set of instantaneous CSIs  and large scale fading gains. However, its complexity is high, hence cannot be readily implemented in dense C-RAN. Recently, we provided a design framework in \cite{pan2017jsac} to design green transmission under partial CSI case. However, the intra-cluster CSI was assumed to be perfect in \cite{Shi2014ICC} and \cite{pan2017jsac}, which is difficult to achieve in practice. Hence, in the paper, we aim to consider the imperfect intra-cluster CSI case, where both the channel estimation procedure for intra-cluster CSI and robust beamforming vector design are considered in the following two sections.

\section{Transmission Scheme Designed for Imperfect Intra-Cluster CSI}\label{robust}

For TDD C-RANs, the training pilots sent from different users to the same RRH should be mutually orthogonal so that the RRH can distinguish the channel vectors from different users. The number of  pilots required scales linearly with the number of users, which becomes excessive for dense C-RANs. Hence, the number of time slots used for data transmission will be significantly reduced. To reduce the number of pilots, they may be reused by a group of users under the condition that none of users in the group shares the same RRH with the other users. It is widely recognized that the pilot reuse scheme will impose the pilot contamination issue, which results in sizeable channel estimation errors.  In the following, we propose a two-stage optimization method to optimize the transmissions for dense C-RAN: In Stage I, a novel pilot reuse scheme is proposed; In Stage II, a robust beamforming-vector optimization algorithm is conceived by considering the pilot contamination incurred by Stage I.

\subsection{Stage I: Novel Pilot Reuse Scheme}\label{stage1}

The pilot reuse issues have been extensively studied in massive multiple-input multiple-output (MIMO) systems \cite{Hyin13}. The basic idea is to reuse the same pilot within the specific group of users having different angles of arrival, which is not applicable in dense C-RAN due to limited number of antennas at each RRH. Recently, Chen \emph{et al.} \cite{Chen2016tvt} proposed a novel pilot allocation scheme for dense C-RANs by using the classic Dsatur graph coloring algorithm, which minimizes the number of pilots required for a given set of users. However, for  simplicity of implementation, the 4G LTE system suggests that the proportion of pilots designated for channel estimation should be fixed within the channel's coherence time, i.e. $1\%$ for a 10 ms training period \cite{Stands}. Given the fixed number of pilots, some users may not be allocated any pilots. Hence, we aim for providing a joint user selection and pilot allocation scheme for maximizing the number of  users admitted at a given number of available pilots, while satisfying the following two conditions:
\begin{enumerate}
  \item The users sharing the same RRH must not reuse the same pilot.
  \item There is an upper-bound on how many time each pilot is reused, denoted as $n_{\rm{max}}$.
\end{enumerate}
The second constraint is imposed for guaranteeing a fair use of the available pilots. In the following, a low-complexity nearly-optimal algorithm is provided to deal with this problem.

\begin{figure}
\centering
\includegraphics[width=4.5in]{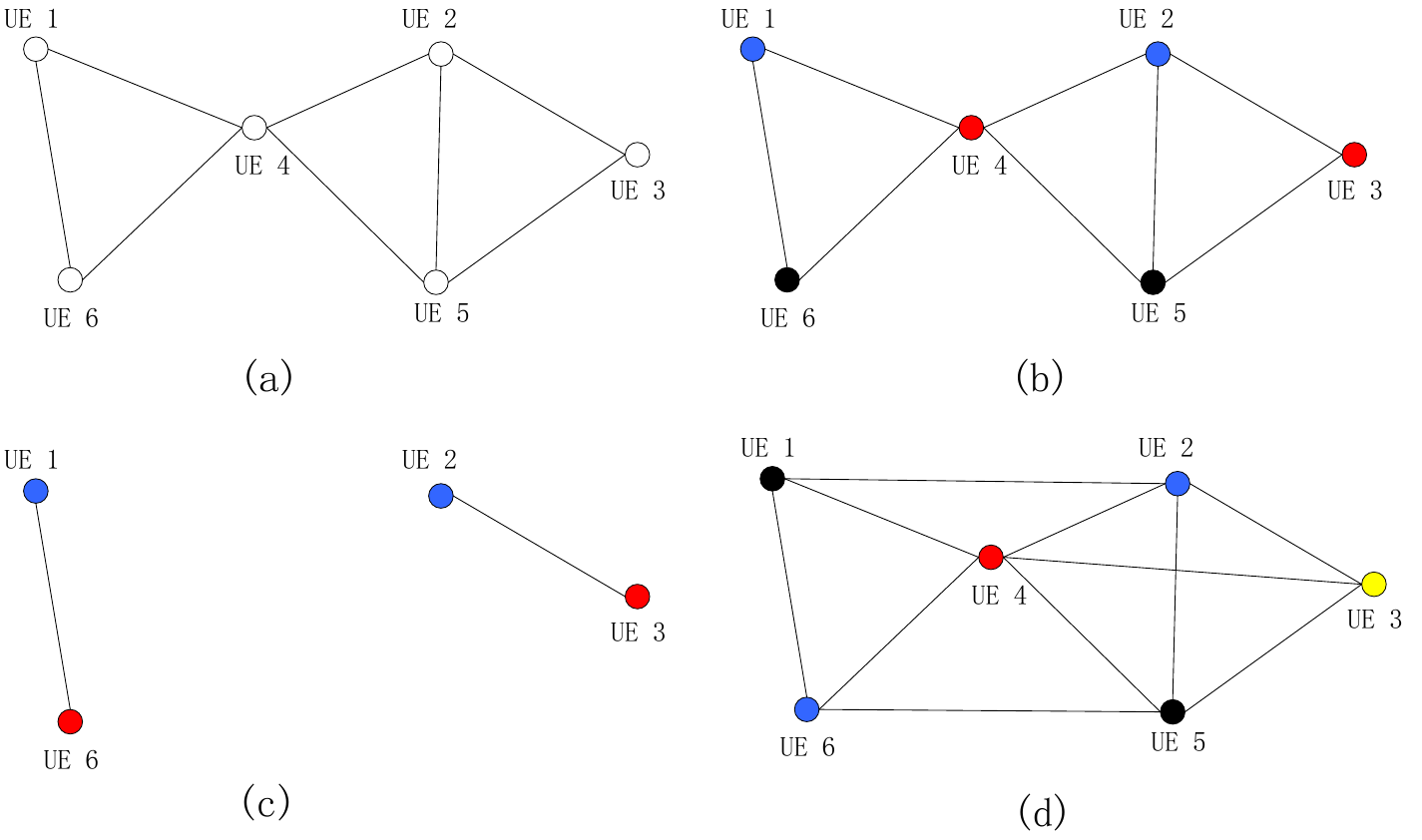}\vspace{-0.5cm}
\caption{(a) Construction of the undirected graph for the network in Fig.~\ref{fig1}-(b); (b) The colored graph after applying the Dsatur algorithm \cite{Chen2016tvt} with $n_{\rm{max}}=2$, the minimum number of required pilots is $n^*=3$; (c) The user selection and pilot allocation result after using the algorithm for Case I when $\tau=2$ and $n_{\rm{max}}=2$; (d) The pilot reallocation result after using the bisection search algorithm for Case II. In this pilot reallocation result, user 1 and use 2 are allocated with different pilots to reduce the pilot interference, and the same holds for user 5 and user 6,  or  UE 3 and UE 4. }\vspace{-0.4cm}
\label{fig2}
\end{figure}

For a dense C-RAN with $K$ users, we construct a $(K\times K)$ matrix ${\bf{B}}$, where each element is given by
\begin{equation}\label{conventionalB}
  {b}_{k,k'} = \left\{ \begin{array}{l}
1,\ {\rm{   if  }}\ {{\cal I}_k} \cap {{\cal I}_{k'}} \ne \emptyset \ {\rm{ and  }}\ k \ne k'\\
0,\ {\rm{  otherwise,}}
\end{array} \right.
\end{equation}
where ${\cal I}_k$ denotes the set of RRHs that potentially serve user $k$.  The above definitions mean that if user $k$ and user $k'$  share common RRHs, the corresponding element is set to ${b}_{k,k'}=1$. Otherwise, the element is  zero. Based on matrix ${\bf{B}}$, a unidirectional graph can be constructed for the network of Fig.~\ref{fig1}-(b)  in  Fig.~\ref{fig2}-(a). Fig.~\ref{fig2}-(b) illustrates the pilot allocation results after applying the Dsatur algorithm \cite{Chen2016tvt}, which shows that at least three different pilots are required.

To solve the above pilot allocation problem, we first adopt the Dsatur algorithm to find the minimum number of pilots required. If this number is higher than the number of  pilots available, some users should be removed. Otherwise, all users can be admitted. In the latter case, some pilots may be reused by up to $n_{\rm{max}}$ users, while some pilots are not allocated, which wastes the pilot resources. To resolve this problem, all available pilots should be used to reduce the pilot contamination. Let us denote the minimum number of pilots required by the Dsatur algorithm as $n^*$. In the following, we provide a detailed algorithm to deal with each case: 1) $n^*> \tau$; 2) $n^*< \tau$. When $n^*= \tau$, the final solution is obtained.

\emph{Case I: $n^*>\tau$}. In this case,  some users should be removed. These users can be rescheduled for transmission in other time slots or frequency bands, which is beyond the scope of this paper. Let us define $\bar{\cal U}$ as the set of all users, and  ${\theta_k}\buildrel \Delta \over = \sum\nolimits_{k' \ne k,k' \in \bar{\cal U}} {{b_{k,k'}}}$ as the degree of the vertex for user $k$ that represents the total number users connected to this user in the constructed undirect graph. The user having the largest value of ${\theta_k}$ should be deleted with high priority, since many users should be allocated different pilots compared to this user. However, there may exist several users with the same value of ${\theta_k}$, and randomly removing one of them will lead to a reduced performance. Hence, the user earmarked for deletion should be carefully selected. Intuitively, the user suffering from the highest pilot interference should be removed. To this end, we  define the metric $\eta_{k,k'}$ to quantify the level of pilot interference between any two users when reusing the same pilot:
\begin{equation}\label{meause}
  {\eta _{k,k'}} = \log \left( {1 + \frac{{\sum\nolimits_{i \in {{\cal I}_{k'}}} {{\alpha _{i,k}}} }}{{\sum\nolimits_{i \in {{\cal I}_k}} {{\alpha _{i,k}}} }}} \right) + \log \left( {1 + \frac{{\sum\nolimits_{i \in {{\cal I}_k}} {{\alpha _{i,k'}}} }}{{\sum\nolimits_{i \in {{\cal I}_{k'}}} {{\alpha _{i,k'}}} }}} \right),
\end{equation}
where $\alpha _{i,k}$ represents the large-scale fading power from RRH $i$ to user $k$. Obviously, a larger  ${\eta _{k,k'}}$ means larger pilot interference between these two users. Then, we quantify the level of pilot interference due to user $k$ by ${\xi _k} = \sum\nolimits_{k' \in {\cal K}_{\pi_k}\backslash \{k\}} {{\eta _{k,k'}}} $ as our figure of merit, where $\pi_k$ denotes the pilot index used by user $k$ and ${\cal K}_{\pi_k}$ represents the set of users that reuses pilot $\pi_k$.
 The user with the largest ${\xi _k}$ should be removed. Hence, we conceive the user selection and pilot allocation method formulated in Algorithm \ref{algorithmcase1}. By invoking it for the network in Fig.~\ref{fig1}-(b), the final user selection result is shown in Fig.~\ref{fig2}-(c).

\begin{algorithm}
\caption{User Selection and Pilot Allocation algorithm for Case I}\label{algorithmcase1}
\begin{algorithmic}[1]
\STATE Initialize matrix $\bf{B}$, user set ${\cal U}$, initial number of required pilots  $n^*$ obtained from the Dsatur algorithm;
\STATE While $n^*  > \tau $
\STATE \qquad Find ${k^*} = \mathop {\arg \max }\nolimits_{k \in {\cal U}} {\theta_k}$. If there are multiple users with the same $\theta_k$, remove  the user with  \\
\qquad   the largest ${\xi _k} $;
\STATE \qquad Remove user $k^*$ from ${\cal U}$, i.e., ${\cal U}{\rm{ = }}{\cal U}/{k^*}$, and update matrix $\bf{B}$ with current ${\cal U}$;
\STATE \qquad Use the  Dsatur algorithm to calculate  $n^*$ with $\bf{B}$ and ${\cal U}$;
\end{algorithmic}
\end{algorithm}

\emph{Case II: $n^*<\tau$.} In this case, we aim for reallocating all the available pilots to all users to additionally reduce the pilot interference. Note that in Fig.~\ref{fig2}-(b), there may be measurable pilot interference between user 1 and user 2. If there are four pilots, these two users can be allocated different pilots as seen in Fig.~\ref{fig2}-(d). We can reconstruct an undirected graph by introducing a threshold $\eta_{\rm{th}}$. If ${\eta _{k,k'}}>\eta_{\rm{th}}$, user $k$ and user $k'$ should be connected.  Based on this idea, matrix $\bf{B}$ can be reconstructed as
\begin{equation}\label{reconsB}
 {b}_{k,k'} = \left\{ \begin{array}{l}
1,\ {\rm{   if  }}\ {{\cal I}_k} \cap {{\cal I}_{k'}} \ne \emptyset \ {\rm{ and  }}\ k \ne k',\\
1,\ {\rm{if}}\  {\eta _{k,k'}} > {\eta _{{\rm{th }}}},\ {{\cal I}_k} \cap {{\cal I}_{k'}} = \emptyset\ {\rm{ and  }}\ k \ne k', \\
0,\ {\rm{  otherwise.}}
\end{array} \right.
\end{equation}
As expected, a smaller value of $\eta_{\rm{th}}$ will result in more users becoming connected with each other and hence more pilots are required. In the extreme case of $\eta_{\rm{th}}<{\rm{min}}\{{\eta}_{k,k'}\}$, all users become connected and the number of  pilots required is equal to $K$. On the other hand, if $\eta_{\rm{th}}\geq {\rm{max}}\{{\eta}_{k,k'}\}$, the reconstructed matrix $\bf{B}$ reduces to the initial $\bf{B}$ defined in (\ref{conventionalB}), where the number of pilots required  is $n^*$. Since $\tau<K$, there must exist a $\eta_{\rm{th}}$ value so that the number of  pilots required is equal to $\tau$. The bisection search algorithm can be adopted to find this threshold $\eta_{\rm{th}}$. Again, Fig.~\ref{fig2}-(d) shows the pilot allocation results after using this algorithm.

\subsection{Stage II: Robust Beamforming-vector Design}\label{stage2}

In Stage II, we aim for designing the beamforming-vectors by considering the pilot contamination due to the pilot reuse scheme in Stage I. Specifically, we formulate a user selection problem for the dense C-RAN where each RRH is equipped with multiple antennas. The following three constraints are considered:
\begin{enumerate}
  \item Each user's data rate should be higher than its minimum requirement;
  \item Each fronthaul link capacity constraint is imposed;
  \item Each RRH has its individual power constraint.
\end{enumerate}
There are three challenges to solve this optimization problem:
\begin{enumerate}
  \item Since we consider the partial CSI case where only the inter-cluster large-scale fading parameters are available, the exact data rate of each user is difficult to obtain.
  \item The fronthaul capacity constraint has the non-smooth and non-differentiable indicator function, which is a mixed-integer non-linear programming (MINLP) problem that is NP-hard to solve \cite{pan2017jsac}.
  \item Due to the channel estimation error, each user suffers from residual self-interference. Conventional weighted minimum mean square error (WMMSE) method of \cite{Qingjiang2011} that has been successfully applied in the  perfect intra-cluster CSI scenario \cite{pan2017jsac} and \cite{Dai2014}, cannot be used for solving the problem here.
\end{enumerate}

We provide brief descriptions of the associated methods to address the above three challenges.

First,  Jensen's inequality is used for finding the lower-bound of the exact data rate, which is more amenable to the design of our algorithm. In \cite{pan2017jsac}, we have shown that for the specific scenario of non-overlapped cluster, the gap between the lower-bound and the exact data rate is within three percent  for both sparse and dense C-RAN scenarios. Hence, a simple correction factor may be used for practical applications.

Second, the indicator function is replaced by a concave  function ${f_\theta }(x) = \frac{x}{{x + \theta }}$, where $\theta$ is a small positive value. The transformed problem is the  difference of convex (d.c.) program, which can be efficiently solved by the successive convex approximation (SCA) method \cite{pan2017jsac}.

Finally, to deal with the final challenge, we adopt the semi-definite relaxation approach and formally prove that semidefinite relaxation is tight with a probability of 1.

\subsection{Simulation Results}\label{simulations}

We now present our simulation results for evaluating the performance of the proposed algorithms. The dense C-RAN is assumed to cover a square shaped area of 700 m $\times$ 700 m. The numbers of RRHs and users are set to $I=36$ and $K=24$ with the densities of 73 RRHs/km$^2$ and 49 users/km$^2$, respectively. This is a typical 5G ultra-dense cellular networks as stated in \cite{xiaohuge2016}. Both the users and RRHs are uniformly and independently distributed in this area. It is assumed that each user is potentially served by
its nearest $L$ RRHs. Each fronthaul link is assumed to only support three users, since mmWave communication is employed as the wireless fronthaul link. The maximum transmit power for each RRH is 100 mW, and the pilot power is 200 mW. The maximum pilot reuse time for each pliot is $n_{\rm{max}}=4$, and the rate requirement for each user is 4 bit/s/Hz.
We compare our proposed algorithm to the following algorithms:
\begin{enumerate}
  \item Orthogonal pilot allocation (with legend ``Ortho"): As the terminology implies, all users are allocated  orthogonal pilot sequences, hence the maximum number of users that can be admitted in Stage I is equal to the number of available pilots $\tau$. These $\tau$ users are randomly selected from $K$ users.
  \item No reallocation operations for Case II in Stage I (with legend ``NoCaseII''): This algorithm is similar to our proposed algorithm, except when Case II would occur, no additional operations are performed.
  \item Conventional pilot allocation method (with legend ``Con''): This algorithm is similar to the ``NoCaseII'' algorithm, except when Case I would occur, the users are randomly removed until the number of required pilots is equal to $\tau$.
  \item Perfect CSI estimation (with legend ``Perfect''): This is the baseline algorithm, where the intra-cluster CSI is assumed to be perfectly known.
\end{enumerate}
\begin{figure}
\centering
\includegraphics[width=3.2in]{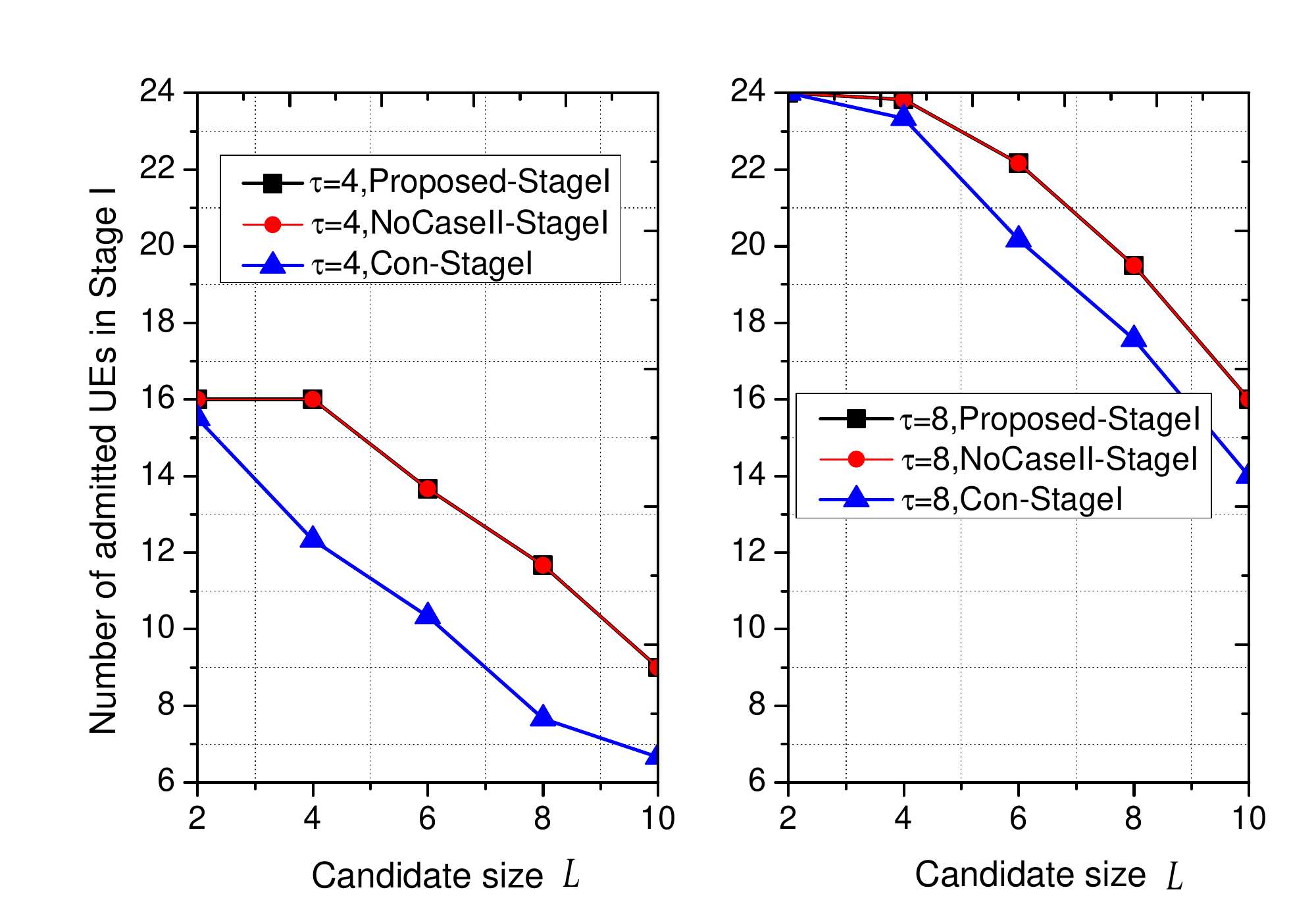}\vspace{-0.2cm}
\caption{Number of admitted UEs in Stage I vs the candidate size $L$. The left subplot corresponds to the case when the number of available pilots is $\tau=4$ while the right one is $\tau=8$.}\vspace{-0.4cm}
\label{fig3}
\end{figure}

\begin{figure}
\centering
\includegraphics[width=3.2in]{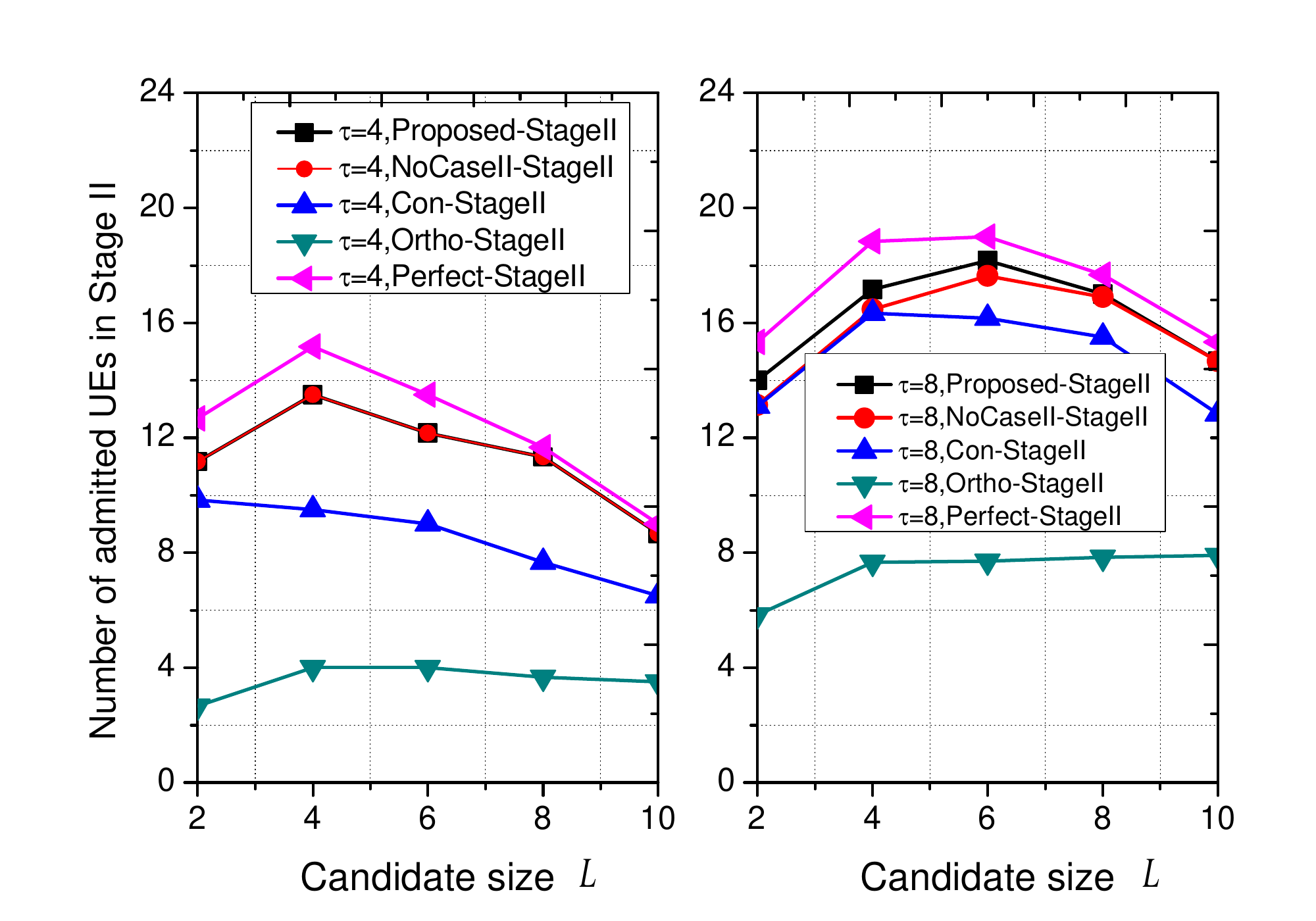}\vspace{-0.2cm}
\caption{Number of admitted UEs in Stage II vs the candidate size $L$.  The left subplot corresponds to the case when the number of available pilots is $\tau=4$ while the right one is $\tau=8$.}\vspace{-0.4cm}
\label{fig4}
\end{figure}

Figures.~\ref{fig3} and~\ref{fig4} illustrate the number of  users admitted vs the candidate RRH set size $L$ in Stage I and Stage II of Section \ref{stage1} and \ref{stage2}, respectively. It is seen from Fig.~\ref{fig3} that the number of admitted users monotonically decreases upon increasing the candidate set size $L$. This is due to the fact that with a larger candidate RRH set size for each user, more users will become connected with each other, which requires more pilots. This figure also illustrates the superior performance of our proposed algorithm over the ``Con'' algorithm, highlighting the necessity of carefully considering the pilot interference, when removing users.

It is interesting to observe from Fig.~\ref{fig4} that the number of  users admitted by all algorithms (except the ``Ortho'' algorithm) initially increases with the candidate set size and then decreases. The reason is that when $L$ starts to increase, a higher spatial degree of freedom is available to support more users. However, when $L$ continues to increase, many users are rejected in Stage I, as seen in Fig.~\ref{fig3}. This trend is different from the widely accepted concept that increasing the candidate set size will always lead to better performance. Hence, the channel estimation process should be taken into account, when designing the cluster. This figure also shows the performance advantage of our proposed algorithms over the other algorithms.

\section{Conclusions and Future Research Challenges}\label{hrforh}

We advocated the application of a user-centric dense C-RAN architecture for UDN due to its appealing features such as the facilitation of centralized signal processing, low hardware cost, etc. However, we also identified the  challenge of requiring a heavy training overhead for estimating all the CSIs for cooperative transmission. As a remedy, we adopted the partial CSI model, where only the large-scale  inter-cluster CSI is available. The channel estimation required for intra-cluster CSI was also considered, where a novel pilot allocation scheme was proposed. Then, we developed a robust transmission design by considering the effect of channel estimation errors. Our simulation results verified the performance advantages of the proposed algorithm over the existing ones.

Finally, we now highlight several promising research directions to make the user-centric dense C-RAN more amenable for practical implementations.

\textbf{Dynamic Cluster Formations}: In this work, the cluster for each user was assumed to be fixed, i.e. each user is only connected to its nearest $L$ RRHs. However, in practical systems, the cluster sizes for users should be adapted to the network state, such as the users' rate requirements or traffic load. Additionally, the channel estimation stage should be taken into account as seen in Fig.~\ref{fig4}, where the network performance may even degrade with the cluster size. How to optimize the cluster size individually for each user by jointly considering the above elements remains an inspiring research direction.

\textbf{User Mobility Management}: User mobility is a very challenging issue in user-centric ultra-dense C-RANs. When the users move from one place to another, the cluster of RRHs assigned for serving this user should be adaptively changed. Explicitly,
 an adaptive mobility management method should be developed so that the serving cluster can follow each user's behavior (e.g. mobility and  service demands) and provide data transmission without the users' involvement. Fortunately, the `Big Data' technique relying on machine learning is becoming mature, which can track the users' mobility and then predict their future locations. By applying this, the users' serving cluster can be formed beforehand that significantly reduces the processing time and meets the targeted quality of experience (QoE) levels.

\section*{ACKNOWLEDGMENT}
This work was supported by the UK Engineering and Physical Sciences Research Council under Grant EP/N029666/1.







\bibliographystyle{IEEEtran}
\bibliography{myre}

\end{document}